\documentclass[twocolumn,twoside,slac_two]{revtex4}
\usepackage{graphicx}
\usepackage{fancyhdr}
\begin{document}

\title{Multiwavelength Opportunities and Challenges in the Era of Public \textit{\textbf{Fermi}} Data}

%

\author{D. J. Thompson}
\affiliation{NASA Goddard Space Flight Center, Greenbelt, MD 20771 USA\\on behalf of the \textit{Fermi} Large Area Telescope Collaboration}
%

\begin{abstract}
The gamma-ray survey of the sky by the {\it Fermi Gamma-ray Space Telescope} offers both opportunities and challenges for multiwavelength and multi-messenger studies. Gamma-ray bursts, pulsars, binary sources, flaring Active Galactic Nuclei, and Galactic transient sources are all phenomena that can best be studied with a wide variety of instruments simultaneously or contemporaneously. Identification of newly-discovered gamma-ray sources is largely a multiwavelength effort. From the gamma-ray side, a principal challenge is the latency from the time of an astrophysical event to the recognition of this event in the data. Obtaining quick and complete multiwavelength coverage of gamma-ray sources can be difficult both in terms of logistics and in terms of generating scientific interest. The {\it Fermi} LAT team continues to welcome cooperative efforts aimed at maximizing the scientific return from the mission through multiwavelength studies.

\end{abstract}

\maketitle

\thispagestyle{fancy}


\section{Opportunities}
During its first year, the \textit{Fermi} Large Area Telescope has excelled in producing scientific results using multiwavelength approaches.  Some examples include:

\begin{itemize}
\item PSR J1741-2054 is a radio pulsar found based on gamma-ray timing \cite{PSR J1741}.  
A bright Fermi LAT point source was the first step.  An analysis of the LAT timing discovered gamma-ray pulsations.  A follow-up observation with the Swift X-Ray Telescope (XRT) found an X-ray source that gave better position information than could be determined from the LAT image.  Using the LAT timing information and the Swift location allowed archival analysis using Parkes radio data and a deep search using the Green Bank Telescope that found the radio pulsar.
\item PMN J0948+0022 is known as a narrow-line quasar or a radio-loud Narrow-Line Seyfert 1 galaxy, a somewhat different class than the blazars that are regularly seen in gamma rays. Contemporaneous observations combining the LAT data with Swift (X-ray, UV, and optical) and Effelsberg (radio) revealed a Spectral Energy Distribution that showed this source to be similar to a blazar, indicating the presence of a relativistic jet \cite{NLS1}.
\item Using the public light curves made available by the LAT team (at the {\it Fermi} Science Support Center Web site \url{http://fermi.gsfc.nasa.gov/ssc/}), Bonning et al. \cite{3C454.3 MW} studied simultaneous multiwavelength variability of blazar 3C454.3 using Small and Moderate Aperture Research Telescope System (SMARTS) telescopes for optical and ultraviolet and X-ray data from the Swift satellite.  They found excellent correlation, with a time lag less than a day, an important parameter for modeling this blazar in terms of an external Compton model.  
\end{itemize}

These few examples illustrate some of the ways {\it Fermi} results enhance scientific understanding when combined with observations and analysis from other wavelengths.  Because the position uncertainties for gamma-ray sources are still large compared to those at many other wavelengths, unidentified gamma-ray sources are inherently subjects for multiwavelength studies, depending on timing, spectral, and modeling to determine what objects produce the gamma rays. 

\section{Enabling Technologies}

Multiwavelength opportunities like those described in the previous section were not possible even a few years ago.  Several developments have facilitated such multiwavelength efforts:

\begin{itemize}
\item Communication - The ubiquity of network connectivity has allowed rapid exchange of data and ideas.  Wireless Internet access and portable devices of all sorts have accelerated the exchange of information.  Campaigns that once had to be organized by telephone and letter can now be arranged in a matter of minutes or hours. 
\item Facilities - Most parts of the electromagnetic spectrum (and several multi-messenger fields) are now covered by ground-based and space-based observatories. {\it Fermi} is just one of many facilities that produce prompt and public results that can be used for multiwavelength study. 
\item Consolidated Information Centers - Resources like ADS, NED, Simbad, ASDC, HEASARC, and others facilitate rapid discoveries of existing coverage of sources.  Scientists can now almost instantaneously review archival results for nearly any cataloged object. 
\end{itemize}


\section{Challenges}
Despite the tools and resources now available for multiwavelength studies, the {\it Fermi} LAT presents some challenges in terms of making the best scientific use of the gamma-ray data.  Three of these issues are described in the sections below. 

\subsection{Challenge 1:Time-Criticality of Response}
Gamma-ray bursts (GRB), thanks to the the Gamma-ray bursts Coordinates Network (GCN) \url{http://gcn.gsfc.nasa.gov/}, offer a paradigm for rapid response to transient astrophysical events.  The success of the GCN originates in part from the intensity of GRB, which can be recognized automatically with high confidence in satellite detectors.  The time for disseminating initial information about a burst is not generally governed by the actual detection but rather by the speed of communication. Multiwavelength studies can often begin within seconds of the initial detection. 

The situation for other gamma-ray sources is more complicated, because none of them approaches the instantaneous brightness of a GRB.  Nevertheless, dramatic changes in flux have been see on time scales of a day or less (e.g. PKS 1502+106, \cite{1502}). On-board analysis of such sources by the LAT is not practical, and so the response depends on both communication and analysis.  The {\it Fermi} LAT data are stored onboard and transmitted through the Tracking and Data Relay System to the ground in batches, not continuously.  This process introduces some delay, as does the data reduction to extract the gamma-ray event candidates and compare the gamma-ray sky with its previous appearance.  Half a day can pass before a flare is discovered.  

The LAT team has taken several steps to minimize the latency in reporting events of multiwavelength interest:

\begin{itemize}
\item Automation - Much of the data handling process is now automated, and efforts continue to streamline the procedures. The analysis pipeline now produces preliminary flux values for over 40 sources of interest routinely, and these results are posted daily at \url{http://fermi.gsfc.nasa.gov/ssc/data/access/lat/msl_lc/}.
\item Dedication - The LAT team has a group of scientists called Flare Advocates or Skywatchers, who examine the automatically produced analysis results as soon as they appear.  By applying scientific expertise at this early stage, the process optimizes the response to findings of astrophysical interest while minimizing any reaction to statistical fluctuations of steady sources.  
\item Communication - Flare advocates use three avenues to share results quickly about activity in the gamma-ray sky.  The first is the use of Astronomer's Telegrams \url{http://www.astronomerstelegram.org/}, over 40 of which have been issued by the LAT team for quickest reporting of results.  The second is a multiwavelength mailing list, gammamw \url{https://lists.nasa.gov/mailman/listinfo/gammamw}, which is used to contact scientists directly about gamma-ray multiwavelength news.  Anyone interested is welcome to join this list.  The third approach is the Fermi Sky Blog, \url{http://fermisky.blogspot.com/}, which posts weekly summaries of the brightest sources in the high-energy gamma-ray sky. 
\end{itemize}

\subsection{Challenge 2: Finding Enough Multiwavelength Coverage}
Although the LAT is an all-sky, every-day monitor for high energy gamma rays, most telescopes at other wavelengths have much smaller fields of view and sky coverage.  In addition, many telescopes have sun-angle constraints.  Multiwavelength coverage of an active gamma-ray source is not assured.  Two approaches are being used by the multiwavelength community to enhance the coverage of the sky:

\begin{itemize}
\item More all-sky or wide-field monitors are becoming available.  The RXTE All-Sky Monitor in X-rays \url{http://xte.mit.edu/} has recently been complemented with the Japanese MAXI all-sky X-ray monitor on the International Space Station \url{http://maxi.riken.jp/top/}. In optical, the Palomar Quest program regularly surveys a large area \url{http://www.astro.caltech.edu/~george/pq/}, and the Pan-STARRS \url{http://pan-starrs.ifa.hawaii.edu/public/} and Skymapper \url{http://msowww.anu.edu.au/skymapper/} programs will be surveying much of the northern and southern hemispheres repeatedly.  At longer wavelengths, Planck \url{http://www.rssd.esa.int/index.php?project=Planck}, Herschel \url{http://herschel.esac.esa.int/}, and WISE \url{http://www.nasa.gov/mission_pages/WISE/main/index.html} are viewing the sky with fairly long cadence.
\item Source monitoring programs have also emerged.  In radio, many observers are cooperating to provide multiple-insrument monitoring of many candidate gamma-ray targets, particularly blazars.  A summary of ongoing activity can be found at \url{http://pulsar.sternwarte.uni-erlangen.de/radiogamma/}.  Similarly, optical programs like the one at the University of Arizona (\url{http://james.as.arizona.edu/~psmith/Fermi/}), SMARTS (\url{http://www.astro.yale.edu/smarts/glast/}), and the GLAST-AGILE Support Program (GASP, \url{http://www.to.astro.it/blazars/webt/gasp/homepage.html})
observe many gamma-ray sources regularly in the optical. 
\end{itemize} 

A useful collection of links to multiwavelength information can be found at \url{https://confluence.slac.stanford.edu/display/GLAMCOG/}.  The LAT team greatly appreciates the ongoing cooperative activities of all these groups and welcomes other telescope teams who participate in particular campaigns. 

\subsection{Challenge 3: Deciding When to Work with the LAT Team}

With all the {\it Fermi} data public, along with software for analysis, anyone can undertake multiwavelength studies incorporating gamma-ray results.  There may be times when contacting the LAT team could benefit such analysis, however.

Analysis of the LAT data does involve some important caveats  (see \url{http://fermi.gsfc.nasa.gov/ssc/data/analysis/LAT_caveats.html} for more details):

\begin{itemize}
\item The diffuse Galactic emission is bright and highly structured. The diffuse model supplied by the LAT team has recently been updated and is likely to continue to evolve.  Separating weaker sources from the diffuse Galactic emission is non-trivial. There are regions of the sky where the diffuse model has deficiencies. 
\item The LAT Instrument Response Functions (IRFs) have significant uncertainties at energies near 100 MeV and a non-negligible charged particle background at energies above 10 GeV. Improvements in the IRFs are expected but are not imminent.  Analysis of data below 100 MeV with the current IRFs is not recommended
\end{itemize}

Some suggestions about when consulting the LAT team might be beneficial:
\begin{itemize}
\item If you are searching for a source that is not in the LAT catalog, then it is probably weak enough that a simple analysis will not be adequate.  
\item If you need a detailed energy spectrum or are looking for particular spectral features, especially at very low or very high energies, the LAT team has experience with non-standard analysis. 
\item If you are trying to analyze the Galactic Center region, you are strongly advised not to go it alone!
\item If you are interested in the most complete multiwavelength coverage, consider contacting the LAT team.  The LAT team knows many cooperating groups across the spectrum who may be interested in working with you (even if you do not include the LAT team).
\end{itemize}

\bigskip 
\begin{acknowledgments}
The \textit{Fermi} LAT Collaboration acknowledges generous ongoing support
from a number of agencies and institutes that have supported both the
development and the operation of the LAT as well as scientific data analysis.
These include the National Aeronautics and Space Administration and the
Department of Energy in the United States, the Commissariat \`a l'Energie Atomique
and the Centre National de la Recherche Scientifique / Institut National de Physique
Nucl\'eaire et de Physique des Particules in France, the Agenzia Spaziale Italiana
and the Istituto Nazionale di Fisica Nucleare in Italy, the Ministry of Education,
Culture, Sports, Science and Technology (MEXT), High Energy Accelerator Research
Organization (KEK) and Japan Aerospace Exploration Agency (JAXA) in Japan, and
the K.~A.~Wallenberg Foundation, the Swedish Research Council and the
Swedish National Space Board in Sweden.

Additional support for science analysis during the operations phase is gratefully
acknowledged from the Istituto Nazionale di Astrofisica in Italy and the Centre National d'\'Etudes Spatiales in France.
\end{acknowledgments}

\bigskip 

\end{document}